\documentclass[runningheads]{llncs}
\usepackage{graphicx}
\usepackage{amssymb}
\usepackage{amsmath}
\usepackage{xcolor}
\usepackage{array}
\usepackage{amssymb}
\usepackage{multirow}
\usepackage{makecell}
\usepackage{soul}
\usepackage{color, xcolor}
\usepackage{float}
\usepackage{cite}

\begin{document}
\title{Residual Moment Loss for \\Medical Image Segmentation}
%\thanks{Supported by organization x.}}
\author{Quanziang Wang\inst{1} \and
Renzhen Wang\inst{1} \and
Yuexiang Li\inst{2} \and
Kai Ma\inst{2} \and
Yefeng Zheng\inst{2} \and
Deyu Meng\inst{1}}
\authorrunning{Q. WANG ET AL.}
\institute{School of Mathematics and Statistics, Xi'an Jiaotong University, Xi'an, China \\
\email{sniperwqza@stu.xjtu.edu.cn} \\
\and
Tencent Jarvis Lab, Shenzhen, China}
%
% \author{Paper ID 2085}
% \institute{Anonymous Organization\\ \email{**@********.***}}
%\maketitle              % typeset the header of the contribution
\maketitle
\renewcommand{\thefootnote}{}
\footnotetext{Q. Wang and R. Wang contribute equally.}
\begin{abstract}
Location information is proven to benefit the deep learning models on capturing the manifold structure of target objects, and accordingly boosts the accuracy of medical image segmentation. However, most existing methods encode the location information in an implicit way, \emph{e.g.,} the distance transform maps, which describe the relative distance from each pixel to the contour boundary, for the network to learn. These implicit approaches do not fully exploit the position information (\emph{i.e.,} absolute location) of targets. In this paper, we propose a novel loss function, namely residual moment (RM) loss, to explicitly embed the location information of segmentation targets during the training of deep learning networks. 
Particularly, motivated by image moments, the segmentation prediction map and ground-truth map are weighted by coordinate information. Then our RM loss encourages the networks to maintain the consistency between the two weighted maps, which promotes the segmentation networks to easily locate the targets and extract manifold-structure-related features. 
% Particularly, the image moments of segmentation prediction and ground truth, indicating the absolute position of objects, are calculated, and our RM loss encourages the networks to maintain the consistency between the two obtained moments, which promotes the segmentation networks to easily locate the targets and extract manifold-structure-related features. 严格来说不是图像矩，相比图像矩少了求和符号
We validate the proposed RM loss by conducting extensive experiments on two publicly available datasets, \emph{i.e.,} 2D optic cup and disk segmentation and 3D left atrial segmentation. The experimental results demonstrate the effectiveness of our RM loss, which significantly boosts the accuracy of segmentation networks.

% Location information is important for locating the ta\maketitlerget objects and capturing their manifold structure in medical image segmentation tasks. Most existing methods commonly encode the relative location information of medical objects by exploiting some correlation between pixels, such as distance. However, these implicit manner do not pay attention to absolute location of pixels, and the segmentation performance and stability is easily affected by pixel values. In this paper, we present residual moment (RM) loss, a novel loss function for training deep segmentation networks. Inspired by image moment, the proposed RM loss explicitly encodes the coordinate information of pixels into the existing architectures. We validate the proposed method by conducting extensive experiments on two public benchmark datasets for both 2D optic cup and optic disk segmentation and 3D left atrial segmentation, and the results demonstrate that our method has a significant performance improvement.

\end{abstract}

%%%%%%%%%%%%%%%%%%%%%%%%%%%%%%%%%%%%%%%%%%%%%%%%%%%%%%%%%%%%%%%%%%%%%%%%%%%%%%%%%
%%%%%%%%%%%%%%%%%%%%%%%%%%%%%%%%%%%%%%%%%%%%%%%%%%%%%%%%%%%%%%%%%%%%%%%%%%%%%%%%%
% Introduction
\section{Introduction}
Image segmentation plays an essential role in the field of medical image analysis, such as disease diagnosis and surgery planning. Witnessing the success of convolutional neural networks (CNNs) in computer vision, an increasing number of researchers developed deep-learning-based frameworks for medical image segmentation \cite{ronneberger2015u,milletari2016v}. However, most of the existing methods simply brought the strategy adopted in computer vision to process medical images, \emph{i.e.,} formulating the segmentation task as pixel-wise classification (\emph{e.g.,} using the cross entropy/Dice loss to supervise the training of segmentation networks). Such a formulation ignores the prior-knowledge---the absolute position information of target objects provides useful information for medical image segmentation, since the to-be-segmented targets often lie in a more prominent low-dimensional manifold \cite{wang2019pairwise}. For example, the same organs of different patients commonly have similar topology structures and positions. Instead, the objects of the same category (\emph{e.g.,} dog) may appear in any position of the natural images with varying sizes. Thus, we uncover a natural question---how to fully exploit the location information to improve the accuracy of medical image segmentation.

% dominates medical image segmentation, due to their promising performance 
% following the same strategy to natural image segmentation
% in natural images may appear in any parts of the images with different size. 

In the recent literature, there are studies embedding location information into deep CNNs for performance improvement, which can be roughly separated into two groups---explicit and implicit embedding. The former one explicitly encoded the geometric position of pixels as features to enhance the capacity of models. For example, Liu \emph{et al.} \cite{liu2018intriguing} proposed CoordConv, which let convolution access to its own input coordinates through concatenating extra coordinate channels. Wang \emph{et al.} \cite{wang2020solo} verified that CoordConv can enable a spatially variant prediction in instance segmentation tasks. Hu \emph{et al.} \cite{2019Local} exploited spatial distances as geometry prior to promote the representation learning of models. Similarly, to learn the local or global geometric dependency, relative geometric position was incorporated into a self-attention module in \cite{bello2019attention,ma2020position}. For medical images, recent studies proposed to use the encoded position information as an additional channel of network input to assist CNNs better dealing with retinal nerve fiber layer defect detection \cite{ding2020retinal} and organ segmentation \cite{murase2020can}. The latter implicit embedding methods usually generated the distance transform maps (e.g., the distance of each pixel to the closest boundary contour) from the ground truth, and designed corresponding loss functions \cite{ma2020distance,kervadec2021boundary,karimi2019reducing,xue2020shape} for optimization.
% , including boundary loss \cite{kervadec2021boundary}, Hausdorff distance loss \cite{karimi2019reducing}, and the signed distance function regression loss \cite{xue2020shape}. 
These methods implicitly encoded the relative position information rather than the absolute one.
% These methods encoded the position information in an implicit way, where the segmentation networks aimed to learn the topology structures of targets embedded into the distance transform maps.

% One is that explicitly
All the above methods verified the benefit yielded by embedding the position information into the deep CNNs to the segmentation performance. In fact, the image moment, which simply embeds the object position with pixel values by element-wise multiplication, is a potential direction to further boost the segmentation accuracy of deep learning networks.
Due to the excellent property of affine invariant \cite{hu1962visual}, image moments have been widely used as a feature extractor in existing works \cite{flusser2006moment,iscan2009moment,zhang2015moment}. Nevertheless, to our best knowledge, no previous study tries to implement the image moment as a loss function for neural networks to directly optimize. In this paper, we propose a novel image-moment-based loss function, termed residual moment (RM) loss. Such a loss function can explicitly embed the coordinate information into the training of segmentation network by calculating the mean squared error of residual map between the prediction map and ground-truth. We mathematically prove that the RM loss is equivalent to a higher order center moment w.r.t. the square of residual map. In essence, this indicates that our RM loss is a specific weighted loss function, where the weights are location-aware and encode the spatial relation of pixels. 
% , which are coupled with the coordinate information.
% In contrast, traditional losses adopted for segmentation task, such as cross entropy and Dice loss, constrain pixel-wise predictions independently, such that the relationship between pixels tends to be ignored. 
In a word, our RM loss can promote the segmentation networks to easily locate the target objects and capture their manifold structure.

In summary, our contributions are mainly three-fold. First, residual moment loss is proposed to make segmentation network capture more position information during training by incorporating coordinate information explicitly. Second, we theoretically prove that the RM loss is equivalent to the higher order center moment of the square of residual map implying that the RM loss is essentially a weighted location-aware loss function. Third, the loss function can be equipped with any existing segmentation networks without changing their architecture and without extra inference burden during testing. We execute comprehensive experiments with both 2D and 3D architectures to verify its effectiveness.

%%%%%%%%%%%%%%%%%%%%%%%%%%%%%%%%%%%%%%%%%%%%%%%%%%%%%%%%%%%%%%%%%%%%%%%%%%%%%%%%%
%%%%%%%%%%%%%%%%%%%%%%%%%%%%%%%%%%%%%%%%%%%%%%%%%%%%%%%%%%%%%%%%%%%%%%%%%%%%%%%%%
% Method
\section{Method}
As aforementioned, our residual moment loss works as an auxiliary loss to help the segmentation networks explicitly capture the location information during training. In the following subsection, we first revisit image moment (Sec. 2.1), and then mathematically formulate the proposed residual moment loss (Sec. 2.2). Finally, we introduce the way to implement the proposed residual moment loss function in a deep learning network (Sec. 2.3). Without loss of generality, we introduce our methods based on 2D images.

\subsection{Revisiting of Image Moment}
Many previous works regard the image moments as statistics to extract some affinity invariant features for the subsequent processing. Mathematically, the two-dimensional $(p+q)$th order moments in statistics are defined as
\begin{equation}
    m_{pq}(f)=\int_{-\infty}^{\infty}\int_{-\infty}^{\infty} i^p j^q f(i,j) di dj, \quad p,q=0,1,2,\cdots,\label{eq:moment}
\end{equation}
where $f(i,j)$ is a bivariate density distribution function of the random variables $i$ and $j$. The center moments are reformulated by replacing $i$ and $j$ with $i-\bar{i}$ and $j-\bar{j}$ in Eq.~\ref{eq:moment}, respectively, where $\bar{i} = m_{10} / m_{00}$ and $\bar{j} = m_{01} / m_{00}$. Due to the sound theoretical properties \cite{hu1962visual}, center moments are widely-used in practice.

Since the random variables $i$ and $j$ are discrete for image processing, the central moments of an $H\times W$ image are rewritten as:
\begin{equation}
    m_{pq}(f)=\sum_{i=1}^{H}\sum_{j=1}^{W} (i-\bar{i})^p (j-\bar{j})^q f(i,j), \quad p,q=0,1,2,\cdots,\label{eq:discrete}
\end{equation}
where $i$ and $j$ represent the pixel position, $f(i,j)$ is the pixel value at position $(i,j)$ of image $f$ and $(\bar{i}, \bar{j})$ is the coordinate of image's center calculated via $\bar{i} = H/2, \quad \bar{j} = W/2$. For simplicity, we marked the $\tilde{i}=i-\bar{i}$ and $\tilde{j}=j-\bar{j}$ in the following. Referring to Eq.~\ref{eq:discrete}, image central moments embed the information of object positions and pixel values by element-wise multiplication.
%, which is robust to affine variant \cite{hu1962visual}.
% Although this encoding method is simple, it is very effective and can construct statistics with affine invariant, such as Hu moments \cite{hu1962visual}.

\subsection{Residual Moment Loss}
In this section, we present the proposed residual moment loss in details. For notation simplicity, we first denote $\mu_{pq}(f)$ as: 
\begin{equation}
    \mu_{pq}(f) = [\tilde{i}^p \tilde{j}^q f(i,j)]_{H \times W}, \quad p,q=0,1,2,\cdots. \label{dense_moment}
\end{equation}

Here, $\mu_{pq}(f)$ is an $H \times W$ matrix with the same size as the input image $f$. Formally, the proposed residual moment loss is defined by the mean squared error (MSE) between $\mu_{pq}$ of the segmentation prediction $\hat{y}$ and the ground-truth $y$, which can be formulated as:
\begin{equation}
    \textit{l}_{RM}^{(pq)} (\hat{y}, y) = \sum_{i=1}^{H} \sum_{j=1}^{W} (\mu_{pq}(\hat{y}) - \mu_{pq}(y))^2.  \label{moment_loss}
\end{equation}

Eq.~\ref{moment_loss} reveals that the segmentation prediction explicitly interacts with the ground-truth not only in terms of object location but also the information of pixel values. Such a flexible information flow enables the segmentation networks to accurately locate the target objects and therefore fully exploit useful information on the manifold structure. Beyond that, we mathematically prove that the proposed loss function is equivalent to the higher order center moment w.r.t. the residual map between the prediction map and the ground-truth map.

% From this equation, we can see that the location information of segmentation targets are explicitly interacted with pixel information, which is helpful for the segmentation networks to locate the target objects and capture their manifold structure during training. Beyond that, we mathematically prove that
% the proposed loss function is equivalent to the higher order center moment w.r.t the residual map between the prediction map and the ground-truth map, i.e.,

\begin{theorem}
\label{theorem}
	Denote the residual map between the prediction map and the ground-truth map as $R=(\hat{y}-y)$. Then
	the loss function Eq.~\ref{moment_loss} is equivalent to the $(2p+2q)$th order center moment of the element-wise square of residual map:
	\begin{equation}
	    \textit{l}_{RM}^{(pq)} (\hat{y}, y) = m_{2p,2q}(R^2). \label{equivalent}
	\end{equation}
\end{theorem}
Henceforth, we name $\textit{l}_{RM}^{(pq)} (\hat{y}, y)$ as \textbf{Residual Moment Loss}.

\begin{proof} Taking the definition of $\mu_{pq}(f)$ into the Eq.~\ref{moment_loss}, we have
    \begin{eqnarray}
	\textit{l}_{RM}^{(pq)} (\hat{y}, y)
	&=& \sum_{i=1}^{H} \sum_{j=1}^{W} (\mu_{pq}(\hat{y}) - \mu_{pq}(y))^2 \nonumber \\
	&=& \sum_{i=1}^{H} \sum_{j=1}^{W} \tilde{i}^{2p} \tilde{j}^{2q} (\hat{y}(i,j) - y(i,j))^2 \nonumber \\
	&=& \sum_{i=1}^{H} \sum_{j=1}^{W} \tilde{i}^{2p} \tilde{j}^{2q} R^2(i,j) = m_{2p,2q}(R^2). \nonumber 
	\end{eqnarray}
The proof is then completed. $\hfill\blacksquare$
\end{proof}

From the proof, we can see that the proposed RM loss actually is the weighted MSE loss, which equals the MSE loss as $p=q=0$. Note that the weights of our RM loss are the coordinates of pixels, which implicitly model the spatial relation between pixels and accordingly make our RM loss location-aware. This is also the underlying reason why the proposed RM loss can help networks to capture the position information naturally. In addition, from the Theorem~\ref{theorem}, the Hu invariant moments \cite{hu1962visual} can also be implemented with the proposed RM loss.

As previously mentioned, our RM loss is an auxiliary loss function assisting the segmentation network. Assuming the commonly used segmentation loss function as $l_{Seg} (\hat{y}, y)$, then, the full objective function can be defined as:
\begin{equation}
\label{total_loss}
    \textit{l}_{total} (\hat{y}, y) = l_{Seg} (\hat{y}, y) + \alpha l_{RM}^{(pq)} (\hat{y}, y),
\end{equation}
where $l_{Seg} (\hat{y}, y)$ can be the widely-used cross entropy loss $l_{CE}$ or Dice loss $l_{Dice}$. In our following experiment, both losses, \emph{i.e.,} $l_{Seg} = l_{CE} + l_{Dice}$, are involved for a better segmentation performance. The hyperparameter $\alpha > 0$ is the weight of $l_{RM}^{(pq)}$. And the choice of order $(p,q)$ is very flexible and we can further use a combination of different orders (verified in Sec. 3.2).

\begin{figure}[t]
\begin{center}
	\includegraphics[width=0.9\textwidth]{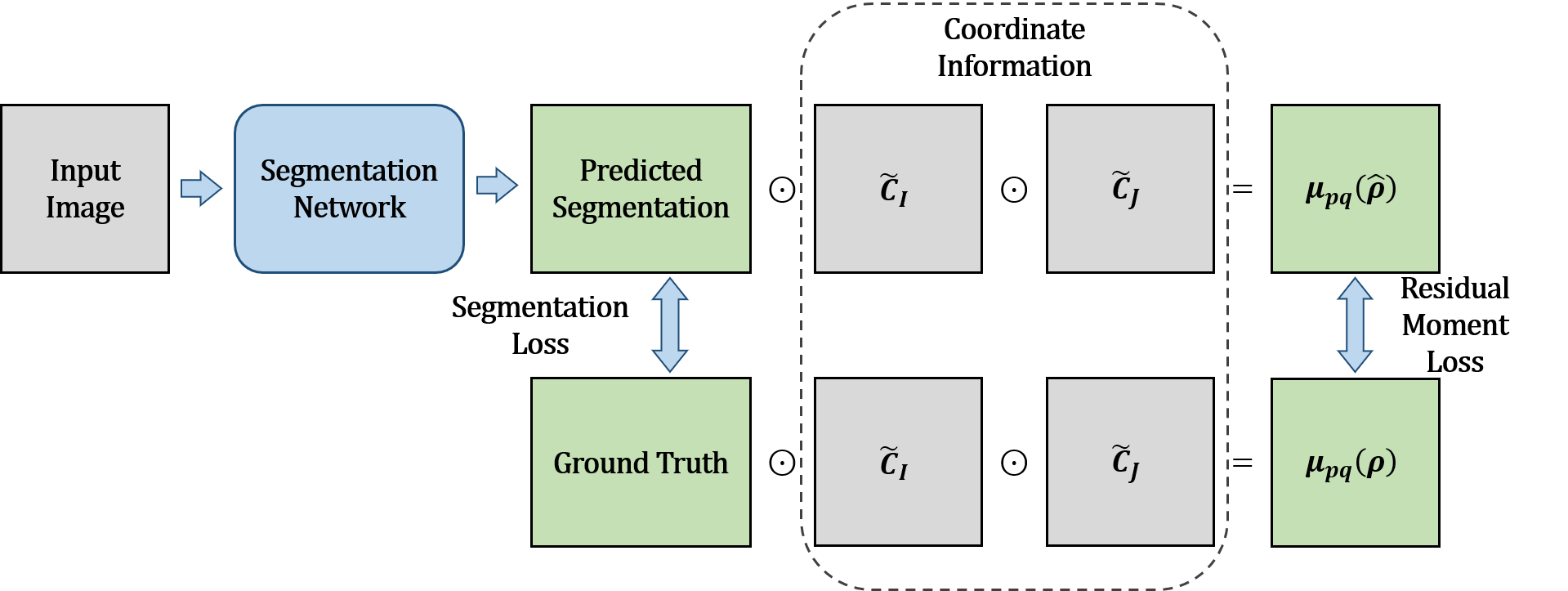}
	\caption{The pipeline of the residual moment loss function. We use the coordinate information to calculate the $\mu_{pq}$. The residual moment loss is the mean squared error (MSE) between the $\mu_{pq}$ of predicted segmentation and ground-truth. And the Segmentation Loss means the common used loss functions, such as cross entropy and dice loss.} \label{network}
\end{center}
%\vspace{-2mm}
\end{figure}

%%%%%%%%%%%%%%%%%%%%%%%%%%%%%%%%%%%%%%%%%%%%%%%%%%%%%%%%%%%%%%%%%%%%%%%%%%%%%%%%%
% This part will introduce how to construct the coordinate information.
\subsection{Coordinate Information}
The whole pipeline of the residual moment loss is shown in Fig.~\ref{network}. Referring to the proof in the previous section, the proposed RM loss adopts the coordinate information as the weights for pixels. 
% However, calculating the residual moment loss with Eq. 4 is not efficient on a graphics processing unit (GPU), which is highly optimized for matrix additions and multiplications.
However, it is wasteful and unnecessary to train the neural network to automatically learn the coordinate information.
Hence, we construct the coordinate matrices ($C_I$ and $C_J$ with the size of $H\times W$) for the coordinate information embedding, which can be represented as:
% The origin coordinate information can be represented as two $H\times W$ matrix
% \begin{equation}
% C_I =
% \begin{bmatrix}
% 1 & 1 & \dots & 1\\
% 2 & 2 & \dots & 2\\
% \vdots & \vdots & & \vdots\\
% H & H & \dots & H
% \end{bmatrix},
% C_J =
% \begin{bmatrix}
% 1 & 2 & \dots & W\\
% 1 & 2 & \dots & W\\
% \vdots & \vdots & & \vdots\\
% 1 & 2 & \dots & W
% \end{bmatrix}.
% \end{equation}
\begin{equation}
    C_I(i,j) = i, \quad C_J(i,j) = j,
\end{equation}
where $i=1,2,\dots,H$ and $j=1,2,\dots,W$. By Eq.~\ref{eq:discrete}, the center coordinate matrices are $\tilde{C}_I = C_I-H/2$ and $\tilde{C}_J = C_J-W/2$. To prevent numerical calculation from being too large, we normalize the coordinate matrices through dividing $\tilde{C}_I$ and $\tilde{C}_J$ by H and W, respectively. Therefore, the $\mu_{pq}$ can be reformulated as:
\begin{equation}
    \mu_{pq}(f) = \tilde{C}_I^p \odot \tilde{C}_J^q \odot f, \quad p,q=0,1,2,\dots,
\end{equation}
where the $\odot$ is the element-wise production. Then we can calculate the RM loss by Eq.~\ref{moment_loss} conveniently.
% The calculation procedure of the coordinate information is shown in Fig.~\ref{network}.
The $\mu_{pq}$ and our RM loss can be extended to 3D by simply switching the coordinate matrices to 3D. The 3D $(p+q+r)$th RM loss is also equivalent to the $(2p+2q+2r)$th center moment of the residual volume.

%%%%%%%%%%%%%%%%%%%%%%%%%%%%%%%%%%%%%%%%%%%%%%%%%%%%%%%%%%%%%%%%%%%%%%%%%%%%%%%%%
%%%%%%%%%%%%%%%%%%%%%%%%%%%%%%%%%%%%%%%%%%%%%%%%%%%%%%%%%%%%%%%%%%%%%%%%%%%%%%%%%
% Experiments and Results
\section{Experiments}
To verify the effectiveness and generalization of our method, we apply the residual moment loss to 2D and 3D neural networks with various public datasets. All our experiments are implemented with the PyTorch (1.3.0) framework \cite{paszke2017pytorch}.

%%%%%%%%%%%%%%%%%%%%%%%%%%%%%%%%%%%%%%%%%%%%%%%%%%%%%%%%%%%%%%%%%%%%%%%%%%%%%%%%%
\subsection{Implementation Details}
We evaluate our method on two datasets, which are the 2D optic cup and disk segmentation dataset: Drishti-GS (GS) \cite{sivaswamy2014GS} and the left atrial (LA) 3D gadolinium-enhanced magnetic resonance imaging (MRI)\footnote{http://atriaseg2018.cardiacatlas.org/} \cite{xiong2021LA}. 

For the GS dataset, we use 50 images for training and 50 for testing. These images are resized to $512 \times 512$ for computational efficiency. We apply the 2D U-Net \cite{ronneberger2015u} as the baseline and set the weight of RM loss in Eq.~\ref{total_loss} as $\alpha=1$. The model is trained by the stochastic gradient descent (SGD) with the learning rate of 0.001 for 2000 iterations. 

The LA dataset contains 100 training cases and 54 testing cases. Following the experimental setting in \cite{ma2020distance}, we randomly selected 16 cases for training and 20 cases for testing for a fair comparison with \cite{ma2020distance}. We cropped all cases centering at the heart region to alleviate the problem of unbalanced categories. All cases are normalized by subtracting the mean and dividing by the standard deviation. We use 3D V-Net \cite{milletari2016v} as the baseline which is optimized by SGD with the 0.01 learning rate. To prevent overfitting, we use dropout during training but turn it off in the inference stage. Besides, the RM loss weight is $\alpha=0.01$ in Eq.~\ref{total_loss}.

The evaluation metrics employed in our experiments include two region-based metrics (\emph{i.e.}, Dice and Jaccard), and two boundary-based metrics (\emph{i.e.}, Average Surface Distance (ASD) and 95\% Hausdorff Distance (95HD)). The region-based metrics and boundary-based metrics can reflect the performance of segmentation and shape similarity, respectively.

%%%%%%%%%%%%%%%%%%%%%%%%%%%%%%%%%%%%%%%%%%%%%%%%%%%%%%%%%%%%%%%%%%%%%%%%%%%%%%%%%
\subsection{Drishti-GS Segmentation Results}
\begin{figure}[t]
	\begin{center}
		\includegraphics[width=0.75\textwidth]{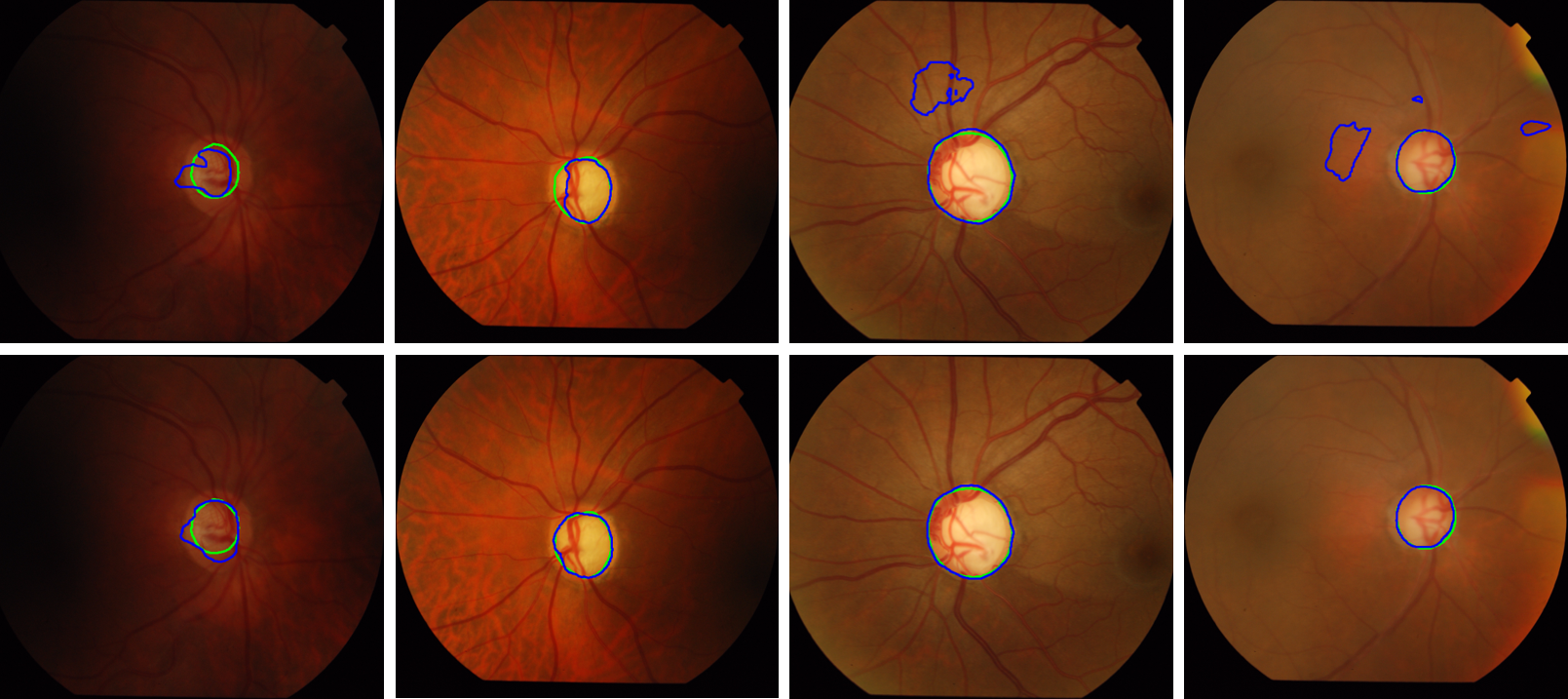}
		\caption{The optic cup and disk results of the GS dataset. The green and blue lines represent the contours of ground-truth and the segmentation results, respectively. The first row is the results of the baseline and the second row is our method, while the first two columns are the cup results and the last two columns are the disk results.} \label{GS_visual}
	\end{center}
	%\vspace{-3mm}
\end{figure}

The quantitative results are listed in Table \ref{GS_result}. We can see that: 1) Our residual moment loss with moments of different orders consistently outperforms the baseline model. 2) Compared to the MSE loss (\emph{i.e.}, the zeroth order RM loss $l_{RM}^{(0,0)}$ with no explicit position information introduced), the proposed RM loss achieves a superior result. 3) The RM loss with a combination of three orders (\emph{i.e.}, (2, 0), (0, 2), and (2, 2)) performs the best on all the metrics. This can be rationally explained as the coordinate information of x and y axes is asymmetric, which leads to the superposition of multiple RM losses of different orders that can introduce various coordinate information. 
From Fig.~\ref{GS_visual}, it can be seen intuitively that our method leads to lower false positive results than the baseline, which is mainly due to its strong capacity of capturing the location information of the cup and disk.

\begin{table}[H]
	\caption{Optical cup and disk segmentation accuracy with mean (standard deviation) on the GS dataset.}\label{GS_result}
	\centering
	\resizebox{\textwidth}{!}
	{
	\begin{tabular}{l|l|p{1.9cm}<{\centering} p{2cm}<{\centering} p{1.9cm}<{\centering} p{1.9cm}<{\centering}}
		\Xhline{0.6pt}
		Dataset & Method & Dice (\%) $\uparrow$ & Jaccard (\%) $\uparrow$ & 95HD $\downarrow$ & ASD $\downarrow$\\
		\Xhline{0.6pt}
		\multirow{2}{*}[-8ex]{GS cup} & U-Net & 80.9 (3.50) & 69.1 (4.66) & 13.9 (1.93) & 6.80 (1.16)\\
		~ & MSE Loss ($l_{RM}^{(0,0)}$) & 81.3 (2.67) & 69.5 (3.46) & 15.9 (2.40) & 7.42 (0.78)\\
		~ & RM Loss $l_{RM}^{(1,0)}$ & 81.3 (2.17) & 69.7 (2.75) & 14.5 (1.69) & 6.81 (0.50)\\
		~ & RM Loss $l_{RM}^{(1,0)}+l_{RM}^{(0,1)}$ & 82.3 (3.94) & 71.0 (5.41) & 15.2 (2.84) & 7.09 (0.93)\\
		~ & RM Loss $l_{RM}^{(2,0)}$ & 83.1 (4.07) & 72.5 (5.46) & 14.4 (2.48) & 6.68 (0.78)\\
		~ & RM Loss $l_{RM}^{(2,0)}+l_{RM}^{(0,2)}$ & 84.0 (2.74) & 73.5 (3.72) & 14.1 (4.12) & 6.72 (1.63)\\
		~ & RM Loss $l_{RM}^{(2,0)}+l_{RM}^{(0,2)}+l_{RM}^{(2,2)}$ & \textbf{84.4 (1.84)} & \textbf{73.7 (1.35)} & \textbf{13.2 (3.28)} & \textbf{6.46 (1.56)}\\
		\hline
		\multirow{2}{*}[-8ex]{GS disk} & U-Net & 92.2 (2.18) & 86.6 (3.28) & 19.7 (15.3) & 7.25 (5.32)\\
		~ & MSE Loss ($l_{RM}^{(0,0)}$) & 91.6 (3.46) & 85.9 (4.96) & 20.2 (16.5) & 7.02 (6.12)\\
		~ & RM Loss $l_{RM}^{(1,0)}$ & 93.0 (2.18) & 88.0 (3.25) & 18.3 (18.01) & 7.14 (6.81)\\
		~ & RM Loss $l_{RM}^{(1,0)}+l_{RM}^{(0,1)}$ & 93.4 (1.29) & 88.5 (2.10) & 18.0 (16.1) & 7.42 (6.94)\\
		~ & RM Loss $l_{RM}^{(2,0)}$ & 93.3 (1.55) & 88.5 (2.04) & 13.0 (2.86) & 4.89 (1.48)\\
		~ & RM Loss $l_{RM}^{(2,0)}+l_{RM}^{(0,2)}$ & 93.0 (1.93) & 87.8 (2.90) & 12.3 (2.67) & 4.60 (1.37)\\
		~ & RM Loss $l_{RM}^{(2,0)}+l_{RM}^{(0,2)}+l_{RM}^{(2,2)}$ & \textbf{94.5 (0.64)} & \textbf{90.1 (0.93)} & \textbf{10.7 (2.24)} & \textbf{3.80 (0.73)}\\
		\Xhline{0.6pt}
	\end{tabular}
	}
	%\vspace{-1mm}
\end{table}

%%%%%%%%%%%%%%%%%%%%%%%%%%%%%%%%%%%%%%%%%%%%%%%%%%%%%%%%%%%%%%%%%%%%%%%%%%%%%%%%%
\subsection{Left Atrial MRI Segmentation Results}
\begin{figure}[t]
	\begin{center}
	\includegraphics[width=0.5\textwidth]{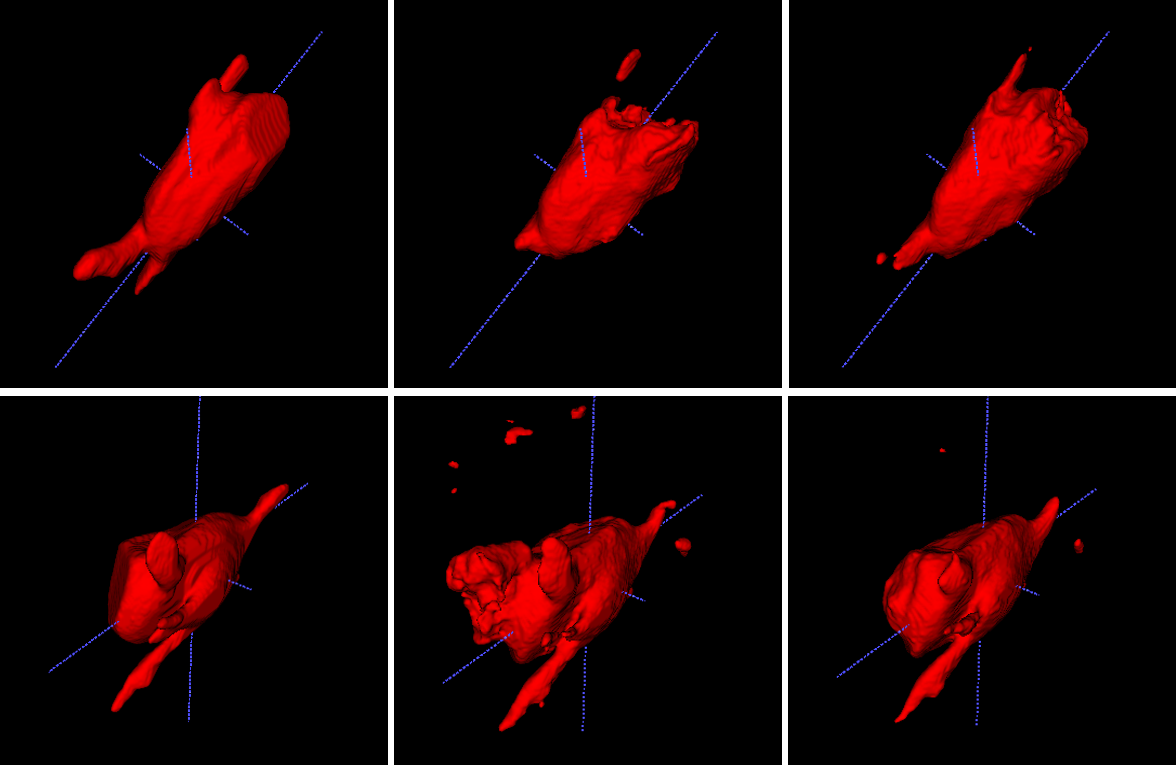}
	\caption{Visualization of the left atrial segmentation results on the LA MRI dataset. The first column is ground-truth. The second and third columns are the results of baseline and our method, respectively.} \label{LA_visual}
	\end{center}
	%\vspace{-1mm}
\end{figure}

Our method can be easily extended to 3D. Table \ref{LA_result} and Fig.~\ref{LA_visual} show the quantitative and qualitative results of the LA dataset. Here, we compare with some implicit position embedding methods, which are boundary loss \cite{kervadec2021boundary}, Hausdorff distance loss \cite{karimi2019reducing} and signed distance function loss \cite{xue2020shape}. The experimental results of these three comparison methods are directly taken from Ma \emph{et al.} \cite{ma2020distance} and our experimental settings followed their work. 

As listed in Table \ref{LA_result}, it can be easily observed that: 1) For Dice and Jaccard, our residual moment loss achieves the best results among all the comparison methods; 2) Our method obtains an obvious improvement on 95HD and ASD compared with the baseline, but fails to outperform implicit position encoding methods for the two boundary-based metrics. This is probably because the loss functions of these methods are specifically designed to minimize the distance-based metrics. As shown in Fig.~\ref{LA_visual}, it is seen that the residual moment loss can significantly improve the ability to extract location information to reduce the misclassification around the objects, which verifies that the proposed loss function is effective in extracting location information.
% According to Table \ref{LA_result}, our residual moment loss has the better Dice and Jaccard. Our method dose not pay much attention to the accurate boundary, that leads to lower 95HD and ASD than boundary loss methods, but it is still better than V-Net baseline. 

\begin{table}[t]
	\caption{Left atrial segmentation accuracy with mean (standard deviation) on the LA MRI dataset.}\label{LA_result}
	\centering
	\resizebox{\textwidth}{!}
	{
	\begin{tabular}{l|p{1.9cm}<{\centering} p{2cm}<{\centering} p{1.9cm}<{\centering} p{1.9cm}<{\centering}}
		\Xhline{0.6pt}
		Method & Dice (\%) $\uparrow$ & Jaccard (\%) $\uparrow$ & 95HD $\downarrow$ & ASD $\downarrow$\\
		\Xhline{0.6pt}
		V-Net & 85.4 (0.69) & 75.2 (1.00) & 26.0 (6.28) & 7.44 (1.81)\\
		\hline
		Boundary Loss \cite{kervadec2021boundary} & 85.0 (5.64) & 74.2 (7.87) & 20.8 (15.0) & 5.43 (3.43)\\
		Hausdorff Distance Loss \cite{karimi2019reducing} & 85.5 (4.96) & 75.0 (7.30) & 15.9 (13.3) & 4.46 (3.68)\\
		Signed Distance Function Loss \cite{xue2020shape} & 84.2 (8.48) & 73.5 (11.0) & \textbf{13.5 (11.2)} & \textbf{3.24 (3.10)}\\
		\hline
		RM Loss $l_{RM}^{(2,0,0)}+l_{RM}^{(0,2,0)}+l_{RM}^{(0,0,2)}$ & 85.9 (0.76) & 75.7 (1.01) & 26.0 (5.67) & 7.13 (1.50)\\
		RM Loss $l_{RM}^{(2,0,0)}+l_{RM}^{(0,2,0)}+l_{RM}^{(0,0,2)}+l_{RM}^{(2,2,2)}$ & \textbf{86.5 (0.51)} & \textbf{76.6 (0.75)} & 21.4 (3.30) & 6.16 (1.01)\\
		\Xhline{0.6pt}
	\end{tabular}
	}
	%\vspace{-1mm}
\end{table}

%%%%%%%%%%%%%%%%%%%%%%%%%%%%%%%%%%%%%%%%%%%%%%%%%%%%%%%%%%%%%%%%%%%%%%%%%%%%%%%%%
%%%%%%%%%%%%%%%%%%%%%%%%%%%%%%%%%%%%%%%%%%%%%%%%%%%%%%%%%%%%%%%%%%%%%%%%%%%%%%%%%
% Conclusion
\section{Conclusion}
In this work, we proposed a novel residual moment loss function to extract location information in medical image segmentation. Motivated by image moments, we explicitly encoded the coordinate information of pixels (or voxels) to the RM loss, which is simple but can capture the target location effectively. In addition, our method is also easy to optimize with high computational efficiency. The experimental results demonstrated that our method could be adapted to various data types and network architectures. The method can also be easily embedded into other network training strategies and used in different practical problems, which will be investigated in our future research. \textit{Source code will be publicly available.}

\bibliographystyle{splncs04}
\bibliography{mybibfile}

\begin{thebibliography}{10}
\providecommand{\url}[1]{\texttt{#1}}
\providecommand{\urlprefix}{URL }
\providecommand{\doi}[1]{https://doi.org/#1}

\bibitem{bello2019attention}
Bello, I., Zoph, B., Vaswani, A., Shlens, J., Le, Q.V.: Attention augmented
  convolutional networks. In: Proceedings of the IEEE/CVF International
  Conference on Computer Vision. pp. 3286--3295 (2019)

\bibitem{ding2020retinal}
Ding, F., Yang, G., Ding, D., Cheng, G.: Retinal nerve fiber layer defect
  detection with position guidance. In: International Conference on Medical
  Image Computing and Computer Assisted Intervention. pp. 745--754. Springer
  (2020)

\bibitem{flusser2006moment}
Flusser, J., Suk, T.: Rotation moment invariants for recognition of symmetric
  objects. IEEE Transactions on Image Processing  \textbf{15}(12),  3784--3790
  (2006)

\bibitem{2019Local}
Hu, H., Zhang, Z., Xie, Z., Lin, S.: Local relation networks for image
  recognition. In: Proceedings of the IEEE/CVF International Conference on
  Computer Vision. pp. 3464--3473 (2019)

\bibitem{hu1962visual}
Hu, M.K.: Visual pattern recognition by moment invariants. IRE Transactions on
  Information Theory  \textbf{8}(2),  179--187 (1962)

\bibitem{iscan2009moment}
Iscan, Z., Y{\"u}ksel, A., Dokur, Z., Kor{\"u}rek, M., {\"O}lmez, T.: Medical
  image segmentation with transform and moment based features and incremental
  supervised neural network. Digital Signal Processing  \textbf{19}(5),
  890--901 (2009)

\bibitem{karimi2019reducing}
Karimi, D., Salcudean, S.E.: Reducing the {Hausdorff} distance in medical image
  segmentation with convolutional neural networks. IEEE Transactions on Medical
  Imaging  \textbf{39}(2),  499--513 (2019)

\bibitem{kervadec2021boundary}
Kervadec, H., Bouchtiba, J., Desrosiers, C., Granger, E., Dolz, J., Ayed, I.B.:
  Boundary loss for highly unbalanced segmentation. Medical Image Analysis
  \textbf{67},  101851 (2021)

\bibitem{liu2018intriguing}
Liu, R., Lehman, J., Molino, P., Such, F.P., Frank, E., Sergeev, A., Yosinski,
  J.: An intriguing failing of convolutional neural networks and the coordconv
  solution. In: Proceedings of the 32nd International Conference on Neural
  Information Processing Systems. pp. 9628--9639 (2018)

\bibitem{ma2020distance}
Ma, J., Wei, Z., Zhang, Y., Wang, Y., Lv, R., Zhu, C., Chen, G., Liu, J., Peng,
  C., Wang, L., et~al.: How distance transform maps boost segmentation {CNNs}:
  an empirical study. In: Medical Imaging with Deep Learning. pp. 479--492.
  PMLR (2020)

\bibitem{ma2020position}
Ma, X., Fu, S.: Position-aware recalibration module: Learning from feature
  semantics and feature position. In: Proceedings of the Twenty-Ninth
  International Joint Conference on Artificial Intelligence (2020)

\bibitem{milletari2016v}
Milletari, F., Navab, N., Ahmadi, S.A.: {V-Net}: Fully convolutional neural
  networks for volumetric medical image segmentation. In: Fourth International
  Conference on 3D Vision. pp. 565--571. IEEE (2016)

\bibitem{murase2020can}
Murase, R., Suganuma, M., Okatani, T.: How can {CNNs} use image position for
  segmentation? arXiv preprint arXiv:2005.03463  (2020)

\bibitem{paszke2017pytorch}
Paszke, A., Gross, S., Chintala, S., Chanan, G., Yang, E., DeVito, Z., Lin, Z.,
  Desmaison, A., Antiga, L., Lerer, A.: Automatic differentiation in pytorch
  (2017)

\bibitem{ronneberger2015u}
Ronneberger, O., Fischer, P., Brox, T.: {U-Net}: Convolutional networks for
  biomedical image segmentation. In: International Conference on Medical Image
  Computing and Computer Assisted Intervention. pp. 234--241. Springer (2015)

\bibitem{sivaswamy2014GS}
Sivaswamy, J., Krishnadas, S., Joshi, G.D., Jain, M., Tabish, A.U.S.:
  {Drishti-GS}: Retinal image dataset for optic nerve head {(ONH)}
  segmentation. In: IEEE 11th International Symposium on Biomedical Imaging.
  pp. 53--56. IEEE (2014)

\bibitem{wang2019pairwise}
Wang, R., Cao, S., Ma, K., Meng, D., Zheng, Y.: Pairwise semantic segmentation
  via conjugate fully convolutional network. In: International Conference on
  Medical Image Computing and Computer Assisted Intervention. pp. 157--165.
  Springer (2019)

\bibitem{wang2020solo}
Wang, X., Kong, T., Shen, C., Jiang, Y., Li, L.: {SOLO}: Segmenting objects by
  locations. In: European Conference on Computer Vision. pp. 649--665. Springer
  (2020)

\bibitem{xiong2021LA}
Xiong, Z., Xia, Q., Hu, Z., Huang, N., Bian, C., Zheng, Y., Vesal, S.,
  Ravikumar, N., Maier, A., Yang, X., et~al.: A global benchmark of algorithms
  for segmenting the left atrium from late gadolinium-enhanced cardiac magnetic
  resonance imaging. Medical Image Analysis  \textbf{67},  101832 (2021)

\bibitem{xue2020shape}
Xue, Y., Tang, H., Qiao, Z., Gong, G., Yin, Y., Qian, Z., Huang, C., Fan, W.,
  Huang, X.: Shape-aware organ segmentation by predicting signed distance maps.
  In: Proceedings of the AAAI Conference on Artificial Intelligence. vol.~34,
  pp. 12565--12572 (2020)

\bibitem{zhang2015moment}
Zhang, Y., Wang, S., Sun, P., Phillips, P.: Pathological brain detection based
  on wavelet entropy and hu moment invariants. Bio-Medical Materials and
  Engineering  \textbf{26}(s1),  S1283--S1290 (2015)

\end{thebibliography}

\end{document}